\documentclass[useAMS,usenatbib]{mn2e}

\usepackage{graphicx}
\usepackage{epsfig}

\newcommand{\ha}{H$\alpha$}

\def\vhel{\ifmmode{V_{{\rm HEL}}}\else{$V_{{\rm HEL}}$}\fi}
\def\vsys{\ifmmode{V_{\rm sys}}\else{$V_{\rm sys}$}\fi}
\def\kms{\ifmmode{~{\rm km\,s}^{-1}}\else{~km~s$^{-1}$}\fi}
\def\vlsr{\ifmmode{v_{\rm lsr}}\else{$v_{\rm lsr}$}\fi}

\def\ltsim{\ifmmode\stackrel{<}{_{\sim}}\else$\stackrel{<}{_{\sim}}$\fi}
\def\gtsim{\ifmmode\stackrel{>}{_{\sim}}\else$\stackrel{>}{_{\sim}}$\fi}

\def\reff@jnl#1{{\rm#1\/}}

\def\aj{\reff@jnl{AJ}}                  
\def\araa{\reff@jnl{ARA\&A}}            
\def\apj{\reff@jnl{ApJ}}                
\def\apjl{\reff@jnl{ApJ}}               
\def\apjs{\reff@jnl{ApJS}}              
\def\ao{\reff@jnl{Appl.Optics}}         
\def\apss{\reff@jnl{Ap\&SS}}            
\def\aap{\reff@jnl{A\&A}}               
\def\aapr{\reff@jnl{A\&A~Rev.}}         
\def\aaps{\reff@jnl{A\&AS}}             
\def\azh{\reff@jnl{AZh}}                        
\def\baas{\reff@jnl{BAAS}}              
\def\jrasc{\reff@jnl{JRASC}}            
\def\memras{\reff@jnl{MmRAS}}           
\def\mnras{\reff@jnl{MNRAS}}            
\def\pra{\reff@jnl{Phys.Rev.A}}         
\def\prb{\reff@jnl{Phys.Rev.B}}         
\def\prc{\reff@jnl{Phys.Rev.C}}         
\def\prd{\reff@jnl{Phys.Rev.D}}         
\def\prl{\reff@jnl{Phys.Rev.Lett}}      
\def\pasp{\reff@jnl{PASP}}              
\def\pasj{\reff@jnl{PASJ}}              
\def\qjras{\reff@jnl{QJRAS}}            
\def\skytel{\reff@jnl{S\&T}}            
\def\solphys{\reff@jnl{Solar~Phys.}}    
\def\sovast{\reff@jnl{Soviet~Ast.}}     
 \def\ssr{\reff@jnl{Space~Sci.Rev.}}     
\def\zap{\reff@jnl{ZAp}}                        
\def\nat{\reff@jnl{Nature}}             

\def\LaTeX{L\kern-.36em\raise.3ex\hbox{a}\kern-.15em
    T\kern-.1667em\lower.7ex\hbox{E}\kern-.125emX}

\begin{document}

\title[VSA measurements of the CMB in an extended array]{High
sensitivity measurements of the CMB power spectrum with the extended Very Small Array}
\author[C. Dickinson et al.]{Clive Dickinson,$^1$\thanks{E-mail:
  cdickins@jb.man.ac.uk}\thanks{Present address: California Institute of Technology, Pasadena, CA 91125} Richard A. Battye,$^1$ Pedro Carreira,$^1$
  Kieran Cleary,$^1$ 
\newauthor Rod D. Davies,$^1$ Richard J. Davis,$^1$ Ricardo
  Genova-Santos,$^3$ Keith Grainge,$^2$ 
\newauthor Carlos M. Guti\'{e}rrez,$^3$ Yaser A. Hafez,$^1$ Michael P.
  Hobson,$^2$ Michael E. Jones,$^2$ 
\newauthor R\"udiger Kneissl,$^2$ Katy Lancaster,$^2$ Anthony Lasenby,$^2$ J. P. Leahy,$^1$ 
\newauthor Klaus Maisinger,$^2$ Carolina \"{O}dman,$^2$ Guy Pooley,$^2$ Nutan Rajguru,$^2$ Rafael Rebolo,$^3$ 
\newauthor Jos\'{e} Alberto Rubi\~no-Martin,$^3$\thanks{Present
address: Max-Planck Institut f\"{u}r Astrophysik,
Karl-Schwazchild-Str.~1, Postfach~1317, 85741, Garching, Germany}
Richard D.E. Saunders,$^2$ Richard S. Savage,$^2$\thanks{Present address: Astronomy Centre, University of Sussex, UK}
\newauthor Anna Scaife,$^2$ Paul F. Scott,$^2$ An\v ze
Slosar,$^{2}$\thanks{Present address: Faculty of Mathematics \& Physics,
University of Ljubljana, 1000 Ljubljana, Slovenia} Pedro Sosa Molina,$^3$
\newauthor Angela C. Taylor,$^2$ David Titterington,$^2$ Elizabeth
  Waldram,$^2$ 
\newauthor Robert A. Watson,$^1$\thanks{Present address: Instituto
de Astrof\'{i}sica de Canarias, 38200 La Laguna, Tenerife, Spain} Althea Wilkinson,$^1$ \\
$^1$Jodrell Bank Observatory, University of Manchester,
  Macclesfield, Cheshire, SK11 9DL, UK.\\
$^2$Astrophysics Group, Cavendish Laboratory, University of
  Cambridge, UK.\\ 
$^3$Instituto de Astrofis\'{i}ca de Canarias, 38200 La
  Laguna, Tenerife, Canary Islands, Spain}

\date{Received **insert**; Accepted **insert**}
       
\pagerange{\pageref{firstpage}--\pageref{lastpage}} 
\pubyear{}

\maketitle
\label{firstpage}

\begin{abstract}
We present deep Ka-band ($\nu \approx 33$~GHz) observations of the CMB
made with the extended Very Small Array (VSA). This configuration produces a
naturally weighted synthesized FWHM beamwidth of $\sim 11$~arcmin
which covers an $\ell$-range of 300 to 1500. On these scales,
foreground extragalactic sources can be a major source of
contamination to the CMB anisotropy. This problem has been alleviated
by identifying sources at 15~GHz with the Ryle Telescope and then monitoring these sources at 33~GHz using a single
baseline interferometer co-located with the VSA. Sources with flux densities $\gtsim 20$~mJy at 33~GHz are subtracted from the data. In addition, we calculate a
statistical correction for the small residual contribution from weaker
sources that are below the detection limit of the survey.

The CMB power spectrum corrected for Galactic foregrounds and
extragalactic point sources is presented. A total $\ell$-range of
$150-1500$ is achieved by combining the complete extended array data
with earlier VSA data in a compact configuration. Our resolution of $\Delta \ell
\approx 60$  allows the first 3 acoustic peaks to be clearly
delineated. The is achieved by using mosaiced observations in 7
regions covering a total area of 82 sq. degrees. There is good
agreement with WMAP data up to $\ell=700$ where WMAP data run out of
resolution. For higher $\ell$-values out to $\ell = 1500$, the
agreement in power spectrum amplitudes with other experiments is also
very good despite differences in frequency and observing technique.
\end{abstract}


\setcounter{figure}{0}

\section{INTRODUCTION}
\label{sec:introduction}

The angular power spectrum of primordial anisotropies in the cosmic
microwave background (CMB) has become an important tool in the era of
``precision cosmology''. Since the first statistical detection of CMB
fluctuations on large angular scales ($\ell =2-30$) by the $COBE$-DMR
instrument (Smoot et al. 1992), several experiments have detected
acoustic peaks in the power spectrum in the $\ell$-range $100-1000$
(Lee et al. 2001; Netterfield et al. 2002; Halverson et al. 2002;
Beno{\^i}t et al. 2003; Scott et al. 2003) and a fall-off in power at
high $\ell$-values  (Dawson et al. 2002; Grainge et al. 2003; Kuo et
al. 2004; Readhead et al. 2004). More recently, the Wilkinson Microwave Anisotropy Probe, henceforth $WMAP$,
has provided unprecedented measurements over the $\ell$-range $2-700$
(Bennett et al. 2003a; Hinshaw et al. 2003a). The WMAP 1-year power
spectrum is cosmic variance limited up to $\ell=350$ and delineates
the first 2 peaks at $\ell \sim 220$ and $\ell \sim 550$ with
exceptional signal-to-noise ratios. The new data have provided detailed
cosmological information on a wide range of parameters (Spergel et
al. 2003) and have raised new questions to be answered. However, the
angular resolution of $WMAP$ limits the power spectrum to $\ell \ltsim
800$. At high $\ell$-values, the origin of the
excess power initially observed by the Cosmic Background Imager (CBI)
at $\ell = 2000-4000$ (Mason et al. 2003) has also generated much
interest. It is clear that the high-$\ell$ CMB power spectrum is one of the
challenges for future CMB experiments including the ESA Planck
satellite (Tauber 2001; Lawrence 2003) due for launch in 2007.
 
The Very Small Array, henceforth the VSA (Watson et al. 2003,
hereafter Paper~I), is a purpose-built radio interferometer that has
measured the CMB angular power spectrum between $\ell = 150$ and 900
in a compact array configuration  (Scott et al. 2003, Paper III) and
more recently up to $\ell=1400$ in an extended array (Grainge et
al. 2003). This paper describes the complete set of extended array
data observed during October 2001 $-$ June 2003. The data cover a larger
area of sky than those analysed by Grainge et al. (2003) by adding
more pointings to the previous VSA fields and by observing 4 new
fields. We use mosaicing techniques to increase the sensitivity,
reduce the sample variance and to facilitate finer $\ell$ resolution
$-$ or equivalently reduce correlations between $\ell$ bins $-$ over
the range $\ell=300-1500$.

The paper is organised as follows. Section~\ref{sec:extended_array}
summarises the telescope parameters and the extended array
configuration. In section~\ref{sec:abs_and_data} we describe the
observations, data reduction and calibration of the data, including a
range of data checks. In section \ref{sec:foregrounds} we discuss the
various foregrounds, particularly discrete radio sources, and the
corrections that were made to the data. The main results, CMB mosaic
maps and CMB power spectrum covering $\ell=150-1500$, are presented in
section~\ref{sec:results}. A morphological analysis of the power
spectrum is given in
section~\ref{sec:ps_gaussians}. Section~\ref{sec:discussion} is a
discussion of the results and comparisons with other data, followed by
conclusions in section~\ref{sec:conclusions}. The cosmological
interpretation of these data, both on their own and combined with
other data, is described in a companion paper (Rebolo et al. 2004).

\section{THE VSA EXTENDED ARRAY}
\label{sec:extended_array}
\begin{figure}
\begin{center}
\includegraphics[width=0.5\textwidth,angle=0]{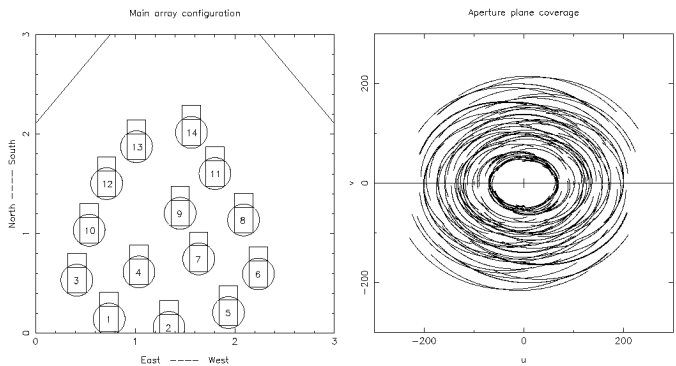}
  \caption{{\it Left}: The extended array configuration of 14 antennas
  on the tip-tilt table. {\it Right}: The corresponding {\it u,v}
  coverage (units of wavelengths) for a 5 hour observation at declination~$+40^{\circ}$. \label{fig:array_config}}
\end{center}
\end{figure}

The VSA is a 14-element interferometer operating in the Ka-band
($26-36$ GHz) situated at the high and dry site of the El Teide
observatory, in Tenerife, at an altitude of 2340~m (see Paper~I). Each
antenna comprises a conical corrugated horn feeding a paraboloid
mirror and mounted anywhere on a $4$-m$\times 3$-m tip-tilt
table located in a metal enclosure to minimise ground emission. Due to
the geometry of the table and enclosure, the VSA  declination range is
restricted to $-5^{\circ}$ to $+60^{\circ}$. In the ``compact''
configuration (Taylor et al. 2003, hereafter Paper~II), the mirrors
were 143-mm in diameter giving a primary beam of $4\degr\!.6$ FWHM at
34~GHz. The ``extended array'', described here, has larger 322-mm
diameter apertures allowing longer baselines while maintaining a high
filling factor and therefore high temperature sensitivity. The longer
baselines increase the angular resolution while the larger horns give
a primary beam of $2\degr\!.1$ FWHM at 33~GHz. This configuration has
a total of 91 baselines with lengths ranging from 0.6~m to 2.5~m as
shown in Fig.~\ref{fig:array_config}. The maximum possible baseline
length, set by the extent of the main tip-tilt table, is considerably
higher with lengths up to $\sim 4$~m. However, for the data presented
here, we chose a limited range of lengths to maximise the overall
temperature sensitivity of the array\footnote{The maximum baseline
lengths possible with the current tip-tilt table, corresponding to
$\ell \sim 2500$, will be utilised in future CMB observations with the
VSA.}. This still allows a wide range of angular scales to be measured
but with well-sampled $u,v$ coverage (Fig.~\ref{fig:array_config}),
which in turn provides good sampling of the power spectrum in $\ell$
and mapping with high fidelity. The filling factor of this
configuration is $\sim 1.6$ times greater than that of the compact
array, leading to an increase in the overall temperature sensitivity
of the array.

The VSA has so far operated in a single channel with instantaneous
band-width of 1.5 GHz. We chose to use the higher end of the band
($\sim 33$~GHz) to minimise foregrounds since the minimum
contamination for total-power measurements is at $\sim 70$~GHz (Banday
et al. 2003; Bennett et al. 2003b). With an average system temperature
of $\sim 35$ K, the VSA achieves an overall instantaneous point source
sensitivity of $\sim 6$ Jy s$^{-1/2}$. This corresponds to a
temperature sensitivity, over a synthesized beam area  ($\Omega_{\rm
synth} \approx 1 \times 10^{-5}$~sr) of $\sim 15$~mK~s$^{-1/2}$. Note
that the exact conversion from flux density to temperature depends on
the beam area and on the $u,v$ coverage which in turns depends on the
declination and the flagging/filtering of the visibility data.  A
typical VSA field at declination $40^{\circ}$  gives a total
$\ell$-range of $\sim 300-1500$ and a naturally weighted synthesized
beam of FWHM $ \sim 11$ arcmin over a $\sim 2^{\circ}.1$
field-of-view. The specifications for the VSA extended array are
summarised in Table~\ref{tab:vsa_specs}.
\begin{table}
\centering
 \caption{Specifications of the VSA in the extended array
 configuration. \label{tab:vsa_specs}}
  \begin{tabular}{lc}
    \hline
Location       			&Izana, Tenerife (2340 m) \\
Latitude			&$+28^{\circ}18^{\rm m}$ \\
Declination range 		&$-5^{\circ}< $Dec $< +60^{\circ}$ \\
No. of antennas (baselines)	&14 (91) \\
Range of baseline lengths	&$0.6$ m $-2.5$ m \\
Centre frequencies		&33, 34~GHz \\
Bandwidth, $\Delta \nu$		&1.5~GHz \\
System temperature, $T_{\rm sys}$ (K)	&$\sim 35$ K \\
Mirror diameters		&322 mm \\
Primary beam			&$2\degr\!.1$ FWHM at 33~GHz \\
Synthesized beam (FWHM)		&$\approx 11$ arcmin \\
Range of $\ell$			&$\sim 300-1500$ \\
Point source flux sensitivity 	& $\sim 6$ Jy s$^{-1/2}$ \\
Temperature sensitivity         & $\sim 15$ mK s$^{-1/2}$ \\ \hline
 \end{tabular}
\end{table}

\section{OBSERVATIONS AND DATA REDUCTION}
\label{sec:abs_and_data}

\subsection{Observations}
\label{sec:obs_schedule}

\begin{figure}
\begin{center}
\includegraphics[width=0.48\textwidth,angle=0]{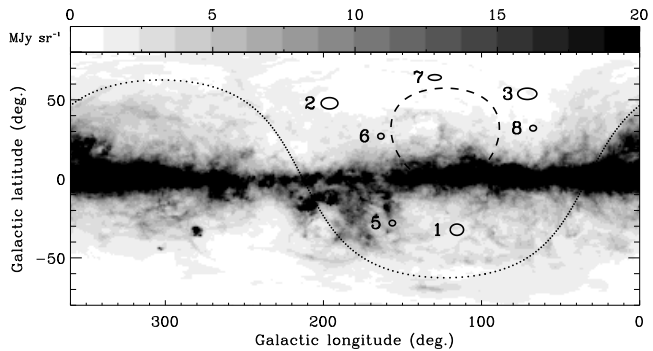}
  \caption{All-sky $100~\mu$m map (SFD98) in Galactic coordinates. The grey-scale is linear
  from $0$ to $20$~MJy~sr$^{-1}$. The locations of the 7 VSA
  regions are shown as circles, stretched in the horizontal direction
  due to the map projection, with diameters scaled up for clarity. The larger ellipses are the 7-field
  regions, labelled 1,2,3 and
  the smaller ellipses, labelled 5,6,7,8, are the 3-field regions. The dashed and dotted
  lines indicate the declination limits for the VSA at $+60\degr$
  and $-5\degr$ respectively.\label{fig:field_positions}}
\end{center}
\end{figure}

The observations presented in this paper were made during the period
October 2001-July 2003 with the array configuration described in
section~\ref{sec:extended_array}. They consist of a total of 33
pointings that make up three 7-field mosaics (VSA1, VSA2, VSA3) and
four 3-field mosaics (VSA5,VSA6,VSA7,VSA8) giving a total area
coverage of 82~sq.~degrees. The sky positions for these 7 regions are
depicted in Fig.~\ref{fig:field_positions} superposed on the
$100~\mu$m all-sky map based on DIRBE/IRAS data (Schlegel, Finkbeiner
\& Davis (1998), henceforth SFD98) in Galactic coordinates. The field
centres and approximate integration times are listed in
Table~\ref{tab:fieldpos_extended}. The total effective integration
time is $\approx 6000$~hours (250 days) after filtering and flagging
of the data. 

The fields were carefully chosen to minimise Galactic and extragalactic emission. This includes avoiding bright
galaxy clusters based on existing catalogues (Ebeling et al. 1998;
Abell 1958) and bright radio sources ($\gtsim 500$ mJy) based on the
NVSS 1.4 GHz survey (Condon et al. 1998) and GB6 survey at 4.85~GHz
(Gregory et al. 1996). Avoiding galaxy clusters is important due to
the potential for Sunyaev-Zeldovich Effect (SZE) decrements to
contaminate VSA data (see section~\ref{sec:sz_contamination}). Diffuse
Galactic emission is minimised by choosing fields with low emission as
predicted by templates of synchrotron, free-free and dust emission
(see section~\ref{sec:gal_foregrounds}); all fields are at Galactic
latitudes $|b| \gtsim 27\degr$. The VSA regions are distributed evenly
in Right Ascension to optimise for 24~hour observations.
Each field is observed on a daily basis for up to a maximum of 5
hours. Where possible, observations are centred on transit to minimise
ground spillover and system temperatures, giving a maximum Hour Angle
range of $\pm 2.5$ hours. Calibrator observations of several bright
radio sources were interleaved between fields for typically 10 min $-$ 1
hour to provide flux and phase calibrations
(section~\ref{sec:calibration}). At the same time as the VSA, a
single-baseline interferometer monitors the brightest sources in the
same patch of sky as part of the source-subtraction process described
in more detail in section~\ref{sec:source_subtraction}.

The results for the first 3 of the 7 fields in the VSA1, VSA2 and VSA3
regions (fields E,F and G) were published earlier (Grainge et
al. 2003; Slosar et al. 2003). In this paper, we extend these regions
(VSA1, VSA2 and VSA3) with a further 4 pointings per region (H,J,K and
L) plus a further 4 new regions (VSA5, VSA6, VSA7 and VSA8) each with
3 pointings (E,F and G). This corresponds to a factor of $\sim 4$ in
the amount of extended array data and a significant increase in
sensitivity over the previous results. Furthermore, the increase in
sky-coverage and mosaicing in each field allows an improvement in
$\ell$-resolution, or equivalently reduced bin-bin correlations, using
mosaicing techniques (section~\ref{sec:results_ps}).
\begin{table}
\centering
 \caption{Celestial (J2000) coordinates for each of the 33 fields
 observed by the VSA in
  the extended array. The effective integration time is calculated after flagging
  and filtering of the data. VSA1, VSA2 and VSA3 are 7-field mosaics and
  VSA5, VSA6, VSA7 and VSA8 are 3-field mosaics giving a total of 33
 pointings.} 
  \begin{tabular}{lccc}
    \hline
       & RA (J2000) & DEC (J2000) & $t_{\rm int}$ (hrs) \\
       \hline
VSA1E 	&00$^{\rm h}$22$^{\rm m}$37$^{\rm s}$	&+30$^{\circ}$16$^{\rm
 m}$38$^{\rm s}$	&126 \\
VSA1F 	&00 16 52	&+30 24 10 	&118 \\
VSA1G 	&00 19 22	&+29 16 39 	&112 \\
VSA1H 	&00 25 04	&+29 09 07 	&142 \\
VSA1J 	&00 13 40	&+29 24 11 	&184 \\
VSA1K 	&00 21 47	&+28 09 08 	&207 \\
VSA1L 	&00 16 09	&+28 16 40 	&315 \\ \hline
VSA2E 	&09 37 57	&+30 41 28 	&131 \\
VSA2F 	&09 43 46	&+30 41 14 	&120 \\
VSA2G 	&09 40 53	&+31 46 21 	&165 \\
VSA2H 	&09 35 00	&+31 46 36 	&308 \\
VSA2J 	&09 46 46	&+31 46 06 	&214 \\
VSA2K 	&09 37 55	&+32 51 28 	&181 \\
VSA2L 	&09 43 52	&+32 51 13 	&253 \\ \hline
VSA3E 	&15 31 43	&+43 49 53 	&172 \\
VSA3F 	&15 38 38	&+43 50 18 	&124 \\
VSA3G 	&15 35 13	&+42 45 05 	&154 \\
VSA3H 	&15 28 25	&+42 44 42 	&213 \\
VSA3J 	&15 42 02	&+42 45 30 	&239 \\
VSA3K 	&15 31 53	&+41 39 54 	&164 \\
VSA3L 	&15 38 35	&+41 40 17 	&332 \\ \hline
VSA5E 	&03 02 57	&+26 11 44 	&196 \\
VSA5F 	&03 08 32	&+26 11 26 	&173 \\
VSA5G 	&03 05 45 	&+27 16 35 	&216 \\ \hline
VSA6E 	&07 28 59	&+53 53 49 	&56 \\ 
VSA6F 	&07 20 30	&+53 54 23 	&141 \\
VSA6G 	&07 24 48	&+55 05 00 	&203 \\ \hline
VSA7E 	&12 32 22	&+52 43 27 	&73 \\ 
VSA7F 	&12 24 07	&+52 43 23	&153 \\
VSA7G 	&12 28 14	&+53 48 25	&206 \\ \hline
VSA8E 	&17 31 34	&+41 57 53	&187 \\
VSA8F 	&17 38 18	&+41 58 22	&300 \\
VSA8G 	&17 34 58	&+40 53 07 	&317 \\   
  \hline
 \end{tabular}
\label{tab:fieldpos_extended} 
\end{table}

\subsection{Data reduction pipeline}
\label{sec:data_reduction}

The data reduction and calibration procedures are similar to those
used for the compact array data (Paper~II). Each observation is
analysed individually using the {\tt reduce} package written
specifically for the VSA. The data reduction procedure is now highly
developed; much of the correcting, flagging, filtering and
re-weighting of the data are performed automatically. This
results in data that are often close to the quality and sensitivity
required for power spectrum estimation. We did however, identify
residual atmospheric contamination which was not fully accounted for
by the standard routines. Such low-level emission can easily be
over-looked once stacked into large data sets with mostly good quality
data. Fortunately, interferometers are much less
sensitive to such signals than total-power instruments, since they
``resolve out'' large angular scale emission due to incomplete $u,v$
coverage at low spatial frequencies. The automatic routines dealt with
correlated emission either by discarding or down-weighting of noisy or
contaminated data. Nevertheless, in a few rare instances, some low-level
contamination still remained. They were initially detected by non-Gaussianity
tests (section~\ref{sec:checks}) which are very sensitive to such
signals. We found that the best way to identify low-level contaminated data was to examine the raw data by eye with no smoothing or calibration applied. Once they were identified, we could remove all residual
atmospheric emission or other non-astronomical signals to well below
the noise on time-scales of a few seconds. The contribution therefore
to the overall data set is negligible. We also searched for noise correlations introduced by such emission. We show that such bias is negligible in
section~\ref{sec:checks}.

We continue to use Fourier filtering to remove the majority of local
correlated signals including the so-called ``spurious signal''
described in Paper~I. The longer baselines and larger
horns of the extended array have significantly reduced this
problem. The filtering removes typically $10-20$ per cent of the
data. The same fringe-rate filtering technique is also applied to the
Sun and Moon. Data are filtered if the Sun and Moon are within
$27\degr$ and $18\degr$ respectively, while if the Sun or Moon are
within $9\degr$ of the field centre then the entire observation is
flagged. No residual Sun or Moon contamination was detected after
stacking the data typically integrated over $50-100$ days. The data
are then further smoothed by a factor of 4 to give 64 sec samples and
a correction (discussed in section~\ref{sec:calibration}) is applied
for the atmospheric opacity which is typically a few per cent at hour angles of  $\sim 2.5$~hours.

The final step for each observation is the re-weighting of the data
based on the r.m.s. noise of each baseline. This is important since
the noise figures can vary substantially between baselines. The
re-weighting also allows the optimum overall noise level to be
achieved. The data are then stacked together either in hour angle, or
in the $u,v$ plane, to reduce the noise level and to facilitate
further flagging of any remaining low-level bad data. Various
consistency checks are performed, to verify the quality of data
reduction; these are discussed in section \ref{sec:checks}. Each VSA field contains $\sim 10^{6}$ visibilities, each of
$64$~sec integration, which are used directly to make maps
(section~\ref{sec:maps}). For power spectrum estimation
(section~\ref{sec:results_ps}), the data are binned in the $u,v$ plane
to reduce the number of data points. Each visibility has an associated
weight calculated by the reduction pipeline by accounting for
integration times, flagging and noise. The reliability of these
weights is verified by comparing them with those calculated from the
data themselves based on the scatter in each $u,v$ cell containing
several independent $64$~sec samples. Both approaches gave results
consistent to within a few per cent. The long-term noise
characteristics of the data are agree very well with the short-term
estimates from individual visibilities.

\subsection{Amplitude and phase calibration}
\label{sec:calibration}

Calibration of VSA data can be split into 4 parts; geometric,
amplitude, phase and atmospheric calibrations. The first stage is to
calculate the exact geometry of the array. This requires knowledge of
the $x,y,z$ positions for each horn accurate to $\ltsim 0.1$~mm. This
also requires calculating the other telescope parameters $-$ including
amplitudes, phases and the mean IF frequency $-$ to converge on the
correct geometry, a total of $\sim 400$ parameters. We use a
maximum-likelihood method developed specifically for interferometric
arrays, introduced by Maisinger et al. (2003). The geometry is
re-calculated from several long observations of Tau-A each time the
array is modified. Thorough testing of the geometry is made by
calibrating observations of other bright radio sources.

Amplitude and phase corrections are calculated from a single
calibrator for each of the 91 complex channels (baselines). An
unresolved, non-variable bright radio source allows the measured
fringes to be corrected for amplitude and
phase. Table~\ref{tab:calibrators} lists the primary calibration
sources used by the VSA with their assumed flux density $S$ (Jy), or
brightness temperature (K), and flux density spectral index, $\alpha$. Note that we adopt the convention $S \propto \nu^{-\alpha}$ for the flux density spectral index. The absolute calibration from correlator units to flux density units (Jy) is described in section~\ref{sec:abs_cal}.
\begin{table}
\caption{Calibrator sources used for amplitude and phase calibration. The absolute temperature scale is tied to a Jupiter brightness temperature of $(146.6
 \pm 2.0)$~K at $33.0$~GHz. Flux density spectral
indices are in the convention $S \propto \nu^{- \alpha}$. \label{tab:calibrators}} 
\centering
  \begin{tabular}{lccc}
    \hline
Source       	&RA/Dec 	&Flux density (Jy) / 	&Spectral  \\
       		&(J2000)	&temperature (K) 	&index $\alpha$ \\	\hline
Tau-A	 	&$05^{\rm h}35^{\rm m}+22^{\circ}01^{\rm m}$	&$333$ Jy (34 GHz)		&$+0.30$ \\
Cas-A		&$23^{\rm h}23^{\rm m}+58^{\circ}49^{\rm m}$   &$168$ Jy (34 GHz)		&$+0.83$ \\
Jupiter		&-		&$146.6$ K (33 GHz)
 &$-2.24$ \\ \hline
\end{tabular}
\end{table}
The increased flux sensitivity of the extended array, compared to the
compact array, means that calibrations are reliable even for short ($<
1$~hour) observations. By cross-calibrating various bright radio
sources with other sources, with no model-fitting, self-calibration or
other post-processing techniques, we found that the majority of VSA
visibility data have phases errors less than  $\sim 10^{\circ}$. We
tested ``secondary'' (antenna-based) corrections based on
calibrations made at the beginning and end of each observation. If
necessary, this can correct for phase drifts during long
observations. However, in general, no correction was required since
the VSA typically remained phase stable for periods of several days as
demonstrated in Paper~I. Uncorrelated phase errors of $\sim 10\degr$ on individual
visibilities have a negligible effect on CMB data since they tend to
average to zero when combining a large number of visibilities. Also, the relatively low signal-to-noise ratios in the
visibility data means that moderate phase errors ($\sim 10\degr$ or
more) can be tolerated since we require a relatively low dynamic range.

Data recorded before 17~July~2002 were taken at a frequency of
$34$~GHz. Due to a technical problem, we replaced the tunable local
oscillator (LO) with a fixed frequency quartz LO. This resulted in a
new centre frequency of $33.0$~GHz for data taken after this
change-over. For the most part, this step in frequency has little
effect on the data as a whole. Nevertheless, fields observed during
the frequency change were reduced and stacked separately to test for
any possible systematic effects introduced by the frequency change. No
significant discrepancies were found (section~\ref{sec:checks}).

Finally, changes in system temperature and in atmospheric opacity with
elevation are corrected, based on the monitoring of modulated noise
signals injected into the VSA system. Each antenna has its own noise
diode to monitor changes in gain or $T_{\rm sys}$ for each
antenna. This is particularly useful for identifying periods of bad
weather and other problems with individual antennae. The correction is
typically a few per cent. Observations with large correction factors
($>20$ per cent), usually caused by bad weather or warm receivers, are
discarded.
\subsection{Absolute flux calibration}
\label{sec:abs_cal}

The absolute flux calibration of VSA data is determined from
observations of Jupiter assuming a brightness temperature, $T_{\rm
Jup}$. Earlier VSA results assumed $T_{\rm Jup} = (152 \pm 5)$~K (3
per cent accuracy in temperature) at 32~GHz, as reported in Mason et
al. (1999). The recent 1-year WMAP data (Bennett et al. 2003a) are
scaled to their own measurement of the CMB dipole with an accuracy of
$0.5$ per cent (Hinshaw et al. 2003b). Page et al.~(2003) have measured Jupiter to
be $T_{\rm Jup} = (146.6 \pm 2.0)$~K at 33.0~GHz
corresponding to an accuracy of $1.5$ per cent in temperature terms,
or equivalently $3$ per cent in the CMB power spectrum ($\Delta
T^{2}$). Note that the WMAP error for $T_{\rm Jup}$ is dominated by
beam shape uncertainties in the WMAP 1-year analysis.

We have used the new WMAP temperature for Jupiter as the
basis for our absolute temperature scale
(Table~\ref{tab:calibrators}).  In principle, this reduces the VSA
calibration error in the power spectrum from 7 to 3 per cent. Indeed,
the new value is consistent with the Mason et al. (1999) value at the $1 \sigma$ level. However, it corresponds to a considerable 8
per cent reduction in the VSA power spectrum values compared to
previous VSA data. With that in mind, we also calculated the flux
density of Tau-A and Cas-A from the 1-year WMAP data. The flux density
values obtained from WMAP alone were found to be $\approx 4$
per cent lower than those using the original VSA value in good
agreement with the new calibration. Earlier VSA power spectrum
measurements should therefore be scaled by a factor of 0.92 to be
consistent with the new calibration scheme used here. A recent paper
describing new CMB results from the CBI (Readhead et al. 2004)
derived a similar scaling factor of $0.94$ for their data, based on
the Jupiter temperature from WMAP.

With angular extents of $\sim5$~arcmin, both Tau-A and Cas-A are
partially resolved by up to $\approx 15$ per cent on
the longest baselines in the extended array. To correct for this, we
used higher resolution maps, made at different wavelengths, as a basis of a model. For Cas-A we used 32~GHz data taken with the 100~m Effelsberg
telescope (W. Reich, priv. comm.). For Tau-A, a 1.4~GHz map made with
the VLA (Bietenholz, Frail \& Hester, 2001) was used and verified with
$850~\mu$m map from the SCUBA instrument (Green 2002). We adopt the
flux density spectral indices given from Mason et al. (1999)
(Table~\ref{tab:calibrators}) to extrapolate between the frequencies. In a few rare instances, we also used  Cyg-A
(33.2~Jy at 34~GHz), Saturn ($T=135.4$~K at 32~GHz) or Venus
($T=442.6$~K at 32~GHz). 


\subsection{Data Checks}
\label{sec:checks}

The VSA data are checked in a number of ways to ensure that the data
are reduced correctly and that there are no systematics or
contamination of the data. The usual checks consist of (i) inspecting
the data by eye, (ii) parallel and independent data reduction, (iii)
stacking the data in different ways, (iv) data splits, (v) simulations
and (vi) a variety of tests to search for residual non-Gaussianity.

The data from each observation are displayed and examined for
contaminating signals such as correlated emission from atmospheric
water clouds unresolved by the telescope beam. The worst periods of
bad weather are easily identified by an increase in system temperature
and overall noise figures. However, unresolved clouds can result in
correlated visibilities without a noticeable increase in system
temperature. We found that in a small portion of the data, correlated
emission remained (section~\ref{sec:data_reduction}). Although such
data made little difference to the final noise level, there is a
concern of introducing noise correlations which could affect the power
spectrum in a biased fashion. Both the reduction software and the
power spectrum estimate software (section~\ref{sec:results_ps}) can
deal with the full noise covariance matrix, but we assume that the
off-diagonal elements were negligible to speed up the calculation of the power spectrum. We tested this assumption by calculating the non-diagonal
elements of the full noise covariance matrix. On typical days the raw
data are consistent with having zero correlations in the noise. We
found that on the worst affected data, the automatic routines
(weighting,filtering, atmospheric calibration and re-weighting)
reduced the correlations to a very large degree. The off-diagonal
elements were found to be negligible in virtually all cases. It is
worth noting, that even if non-zero off-diagonal terms remain, they
will tend to average away to zero due their complex (real and
imaginary) nature, as more days of data are combined. A further test
was to compare the r.m.s. level well outside the primary beam with
that computed for an ``auto-subtracted'' map (which is free from CMB
or any other correlated signal; see Paper~II). We found that the noise
is indeed uncorrelated from visibility to visibility which assures us
that the noise really is decreasing with the square root of time. 

The reduced data are stacked and combined in a variety of ways to look
for possible contamination or other instrumental affects. By stacking
the complete data set in Hour Angle, for a given field, any spurious
signals (see Paper~II) that had not been filtered out, would add
coherently for each baseline and were therefore easy to detect and
flag. For the extended array, very few instances of this effect were
seen. In addition, the real and imaginary components of the visibility
stacks were inspected in the $u,v$ plane.

We note that where possible, the data are reduced in parallel, by at
least two of the three institutions in the VSA collaboration, in order
to perform consistency checking. We compared the fields analysed
independently by different institutions to check that there were no
major data reduction issues due to variations in cut-off levels,
filtering parameters and subjective flagging of the data. No
significant differences were found. The same comparisons were made in
the power spectra which were also found to be completely consistent
with each other.

As an additional check on data consistency, the data for each VSA
field were split by epoch and the $\chi^{2}$ statistic was computed. Since the sky signal is the same in both halves of the data, the $\chi^{2}$ statistic can be used
to test for the presence of systematic errors. Table~\ref{chisq.table} gives the $\chi^{2}$ values and
associated significances. The term significance in this context
represents the probability of exceeding the observed $\chi^{2}$ value
in the cumulative distribution function. For each field, and averaged
over all fields, the reduced $\chi^{2}$ is $\sim 1$ with a combined significance of $0.50$. This indicates that the splits are statistically consistent with each other.
\begin{table}
\centering
\caption{The $\chi^{2}$ values for data splits on each of the VSA fields. In each case the visibility data from each field were split in two according to epoch and the $\chi^{2}$ of the difference vectors formed. Also tabulated are the number of degrees of freedom (d.o.f.) and the significance of each $\chi^{2}$ value (see text). Fields which straddle the observing frequency change-over date are separated into 34~GHz (`A') and 33~GHz (`B') stacks.}
\label{chisq.table}
\begin{tabular}{lccc}
\hline
Field  & d.o.f. & $\chi^{2}$ & Significance\\
\hline
VSA1E&4778&4851&0.23\\
VSA1F&3871&3964&0.14\\
VSA1G&4070&3752&0.99\\
VSA1H\_A&3867&3790&0.81\\
VSA1H\_B&3655&3800&0.04\\
VSA1J&3845&3890&0.30\\
VSA1K&3706&3691&0.56\\
VSA1L&3742&3733&0.54\\
VSA2E&5002&5170&0.05\\
VSA2F&3831&3763&0.78\\
VSA2G&4314&4297&0.58\\
VSA2H&4142&4052&0.84\\
VSA2J\_A&4120&4161&0.32\\
VSA2J\_B&3963&4142&0.02\\
VSA2K&3827&3731&0.86\\
VSA2L&3820&3996&0.02\\
VSA3E&4937&5011&0.22\\
VSA3F&4300&4325&0.39\\
VSA3G&4970&5060&0.18\\
VSA3H\_A&4236&4252&0.42\\
VSA3H\_B&3926&3892&0.64\\
VSA3J&4105&4081&0.60\\
VSA3K&3965&4009&0.31\\
VSA3L&4354&4338&0.56\\
VSA5E&3692&3427&0.99\\
VSA5F&1627&1651&0.33\\
VSA5G&3271&4762&0.00\\
VSA6E\_A&701&669&0.80\\
VSA6E\_B&1297&1279&0.63\\
VSA6F&3578&3507&0.80\\
VSA6G&2085&2119&0.29\\
VSA7E\_A&711&716&0.43\\
VSA7E\_B&2623&2572&0.75\\
VSA7F&3560&3532&0.63\\
VSA7G&3625&3639&0.43\\
VSA8E\_A&3039&3015&0.62\\
VSA8E\_B&1827&1796&0.69\\
VSA8F&2003&1986&0.60\\
VSA8G&3983&4064&0.18\\
\hline
 & &~~Mean~$=$ &0.50 \\ \hline
\end{tabular}
\end{table}
A wide range of non-Gaussianity tests were performed on individual
observations and on the stacks. The techniques and results from
earlier analyses are presented in Savage et al. (2004) and Smith et
al. (2004). Detections of non-Gaussianity in the final stacked data
were traced to a small portion of the data and hence cannot be
cosmological in origin. These data were excised before calculating the
final power spectrum in section~\ref{sec:results_ps}.


\section{Foregrounds}
\label{sec:foregrounds}

\subsection{Source subtraction}
\label{sec:source_subtraction}

\subsubsection{Observational strategy}

Discrete sources are the largest foreground for the VSA. The power
spectrum of Poisson-distributed sources increases as $\ell^{2}$
(Taylor et al. 2001) while that of the CMB decreases exponentially
with increasing $\ell$. The point-source
contribution will therefore dominate at higher $\ell$-values in the
absence of an effective source-subtraction strategy. We carefully
choose regions of the sky to not contain bright ($\gtsim 500$~mJy at
1.4~GHz) radio sources (section \ref{sec:obs_schedule}) when
extrapolated to 33~GHz from NVSS (1.4~GHz) and GB6 (4.85~GHz)
catalogues. However, the contamination from sources below $\sim
500$~mJy is still significant, particularly at the highest $\ell$-values.

To correct for any sources unidentified in the above surveys, the VSA
employs a unique two-stage source subtraction strategy. The first
stage is to survey the regions observed by the VSA with the Ryle
Telescope at 15~GHz to a limiting flux density of $\sim 10$~mJy
(Waldram et al. 2003). This limit was chosen primarily to make sure
that all sources that could potentially contaminate VSA data are
located, even if they have rising spectra, $\alpha < 0$, between 15
and 30~GHz. Only the very steepest inverted spectrum ($\alpha < -1$)
sources will be missed using this technique. The results from the
recent 9C survey (fig.~9 in Waldram et al. 2003), from the Ryle
Telescope at 15~GHz and a completeness limit of 25~mJy, indicate that
less than 1 per cent of their radio sources have $\alpha < -1$. We
therefore expect little or no effect from such a rare population of
sources at these flux density levels. The second stage, is to
follow-up these sources with a single-baseline interferometer at the
same frequency as the VSA {\em and} at the same epoch; each source is
observed many times with an interval of a few days. The source
subtraction interferometer consists of two 3.7~m dishes with a 9~m N-S
separation corresponding to a resolution of $\sim 3$ arcmin and has a
sensitivity of $\approx 340$~mJy~s$^{-1/2}$. The dishes are located in
separate identical enclosures to minimise ground spill-over and
cross-talk.

\subsubsection{Source subtractor observations and data reduction}

For a typical observation, the drive system of the source subtractor
takes the coordinates of the previously identified 15~GHz sources
within a radius of $\sim 2^{\circ}$ of the VSA field centre. A
different subset (typically $\sim 30$) of these sources is monitored
over the course of each $\sim 5$-hour observing run. In this manner,
the flux density of the sources is sampled over the course of the
observations of a given field. This takes into account source
variability on time-scales longer than a few days.

The single baseline interferometer has identical back-end hardware to those of
the main array and the data are processed in a similar way. The
primary flux calibrator for the source subtractor is the planetary
nebula NGC~7027. We assume a flux density of ($5.45 \pm 0.20$)~Jy at
32.0~GHz and a spectral index $\alpha = (0.1 \pm 0.1)$ (Mason et
al.~1999), corresponding to a $\sim 4$ per cent calibration
uncertainty. Variability of NGC~7027 is below 3 per cent (Peng et
al.~2000) at VSA frequencies, apart from a secular decrease with time
of $\sim 0.6$ per cent per year (Ott et al.~1994). Phase calibration
is provided by interleaved observations of the brightest sources
($\gtsim 100$~mJy) in each field.

\subsubsection{Results}
\label{sec:source_results}

The result of the source monitoring programme is
two-fold. Firstly, the 33~GHz flux densities of the sources in the VSA fields
are known (at the time of the main array observation)  and are
subtracted from the visibility data down to a level of 20~mJy. The details of the complete source survey
will be presented by Cleary et al. (in prep.).  Secondly, the source
counts derived from the observations can be used to apply a
statistical correction due to fainter sources below the subtraction
limit.  This analysis assumes that the contribution from clustered
sources is negligible compared to that from the Poisson
component. This is a reasonable assumption on the basis of estimates
for the Planck 30~GHz channel by Toffolatti et al. (1998).

Source monitoring for VSA source subtraction was performed at 34~GHz
and 33~GHz. In order to derive the source counts, the flux density of
each source measured at 34~GHz was corrected to 33 GHz using its
$\alpha_{15.2}^{34}$ spectral index. An average was then taken of the
source measurements over the entire period of extended array
observations.  Of the 453 sources monitored, 131 were found to be $>
20$~mJy at 33~GHz.

The 33~GHz source counts at flux densities less than $\sim 5$~mJy are
unknown; however the counts are expected to flatten with decreasing
flux density. Hence, simply extrapolating the measured VSA source
counts to fainter fluxes would over-estimate the residual source
contribution. In the absence of low flux density data, the
measured VSA source counts were used to re-scale the Toffolatti et
al. (1998) 33 GHz differential source count model. Integrating the
rescaled model up to 20 mJy, we get an estimate of the contribution
from faint sources below the source subtraction limit. The residual
source power spectrum ($\sim 210~\mu{\rm K}^{2}$ at $\ell =1000$) is
then binned using the VSA window functions and directly subtracted
from the band-power estimates as an uncorrelated statistical
correction. The correction applied to these data is $\Delta T_{\rm
src}^{2} = 210 \times (\ell / 1000)^{2}$ in units of $\mu$K$^{2}$.

\subsection{Galactic foregrounds}
\label{sec:gal_foregrounds}

\subsubsection{Overview}

At frequencies of $\sim 30$~GHz, particularly at large angular scales
($\ell < 100$), emission from the Galaxy can contaminate CMB data
(Banday et al.~2003;~Bennett et al.~2003b). There are currently three
well-established diffuse Galactic foregrounds: synchrotron emission
from relativistic electrons spiralling in the Galactic magnetic field,
free-free (thermal bremsstrahlung) emission from ionized gas, and
vibrational dust emission. The power spectrum of Galactic emission
falls with increasing $\ell$ (for example, see Giardino et al.~2001)
and hence it is not expected to be a major contaminant at smaller
angular scales. The VSA is insensitive to large angular scales
($\gtsim 1\degr$) due to the incomplete $u,v$ coverage and primary
beam attenuation. Furthermore, the VSA fields have been carefully
selected to be located in ``cold'' areas of sky
(section~\ref{sec:obs_schedule}), at high Galactic latitudes, thus
minimising any potential contamination from the Galaxy.  There is
however, evidence for another foreground component, recently nicknamed
``foreground X'' (de Oliveira-Costa et al.~2004), that has been shown
to be strongly correlated with FIR emission ($\lambda \sim 100~\mu$m)
in the frequency range $\sim 10-50$~GHz (Kogut et al.~1996; Leitch et
al.~1997; de Oliveira-Costa et al.~2002; Mukherjee et al.~2002; Banday
et al.~2003; Bennett et al.~2003b). A recent discussion on the
possible origins of the anomalous emission, including spinning dust
(Draine \& Lazarian, 1998) and magnetic dust grains (Draine \&
Lazarian, 1999) is given by Banday et al. (2003). The 1-year WMAP data
also detect the dust-correlated component in the K, Ka and Q bands
(Bennett et al.~2003b). Using the statistical distribution of derived
spectral indices, they interpret this to be synchrotron emission with
a relatively flat spectral index, $\beta \approx 2.5$ ($T \propto
\nu^{-\beta}$), and constrain the spinning dust component to
contribute less than 5 per cent to the foregrounds. They argue that
the 408~MHz map is not a good template for synchrotron emission at
higher frequencies ($\gtsim 20$~GHz) because it is dominated by steep
spectrum synchrotron emission with $\beta \approx 2.9$. This is a
contentious issue which has already been re-examined but with no
consensus on the origin of the anomalous component (e.g. Lagache 2003;
Banday et al.~2003; Finkbeiner et al.~2002; Finkbeiner 2003a; Casassus
et al.~2004; de Oliveira-Costa et al.~2004). There is nonetheless
clear evidence for a dust-correlated component, at frequencies of
$\sim 30$~GHz, with a typical coupling coefficient to the SFD98
$100~\mu$m map of $T_{\rm b}/I_{\rm 100} \sim
10~\mu$K/(MJy~sr$^{-1}$). Indeed, the VSA has observed selected dust
clouds to further investigate the Galactic emissions at 33~GHz
(Dickinson et al., in prep.).

\subsubsection{Galactic foreground estimates}
\label{sec:diffuse_foreground_estimates}

To quantify the Galactic foreground we use external templates since a
spectral analysis is not possible for single frequency data. The
template maps are (i)~the 408~MHz all-sky map at a resolution of
51~arcmin (Haslam et al.~1981) for steep spectrum synchrotron,
(ii)~\ha~data from the Wisconsin H-Alpha Mapper (WHAM) at $1\degr$
resolution (Haffner et al.~2003) for free-free emission and (iii) the
$100~\mu$m map at a resolution of 6.1~arcmin (SFD98) for
dust-correlated emission. The synchrotron and free-free templates do
not have adequate resolution to permit a full cross-correlation
analysis, although high resolution \ha~data will be available in the
near future (Dennison, priv. comm.) from the Virginia-Tech
Spectral-line Survey (VTSS; Dennison,~Simonetti \& Topasna 1998). We
therefore calculate the r.m.s. fluctuations in each region at the
resolution of $1\degr$ in a similar manner to the analysis used in
Paper~II using simple foreground models to convert to temperature
fluctuations at 33~GHz. We use circular regions over a $4^{\circ}\!.5$
and  $3^{\circ}\!.0$ diameter for the 7-field mosaics and 3-field
mosaics respectively. We assume a spectral index for synchrotron of
$\beta = 2.9$ and a conversion factor from \ha~intensity, in units of Rayleigh (where $1 R \equiv 10^{6}/4\pi$ photons s$^{-1}$~m$^{-2}$~sr$^{-1}$), to brightness temperature at 33.0~GHz, of $5.0~\mu$K$~R^{-1}$ (Dickinson, Davies \& Davis 2003; Finkbeiner 2003b). The r.m.s. values were negligible
($\ltsim 10~\mu$K$^{2}$) in the VSA fields compared to the CMB
fluctuations ($\gtsim~1000 ~\mu$K$^{2}$). 

For the dust-correlated component, for which we have a high resolution
(6.1~arcmin) template, we smoothed the SFD98 $100~\mu$m map to
22~arcmin ($\ell \sim 1000$) and estimated the foreground levels by
assuming a typical coupling coefficient between brightness temperature at
33~GHz and the $100~\mu$m intensity of $T_{\rm b}/I_{100} =
10~\mu$K/(MJy~sr$^{-1}$). The r.m.s. estimates are listed in
Table~\ref{tab:gal_foregrounds}. At $\ell=1000$, the CMB fluctuations are at
$\Delta T_{\rm rms}^{2} \sim 1000~\mu$K$^{2}$ while the
r.m.s. foreground estimations are $\ltsim 10~\mu$K$^{2}$. Hence, for most
of the VSA regions, the Galactic emission is essentially negligible
unless they have an unusually large coupling coefficient ($\gtsim
30~\mu$K/(MJy~sr$^{-1}$)). From the r.m.s. estimates in
Table~\ref{tab:gal_foregrounds}, the VSA1 and VSA5 regions seem
to have stronger Galactic emission compared to the
other fields. Nevertheless, these are still below the CMB
anisotropies. Furthermore, there is no evidence of peculiarities when
comparing these fields with the other VSA fields both in terms of the
reconstructed maps (section~\ref{sec:maps}) or the power spectra of
individual VSA regions discussed further in
section~\ref{sec:discussion}.
\begin{table}
\centering
 \caption{Galactic dust-correlated foreground power estimates (r.m.s)
 at $\ell = 1000$ for each VSA region based on the SFD98 $100~\mu$m map. We assume a
 correlation coefficient, between the foreground map (in units of MJy~sr$^{-1}$) and 33~GHz brightness temperature (in units of $\mu$K), of $10~\mu$K/(MJy~sr$^{-1}$).}
  \begin{tabular}{lc}
    \hline
VSA region	& r.m.s. power fluctuations	\\ 
		& ($\mu$K$^{2}$)	\\ \hline
VSA 1 		& 31.9			\\	
VSA 2		& 1.3		\\
VSA 3 		& 7.9			\\ \hline	
VSA 5		& 88.4			\\
VSA 6 		& 9.6			\\	
VSA 7		& 2.3			\\
VSA 8 		& 9.0			\\ \hline	
\label{tab:gal_foregrounds} 
 \end{tabular}
\end{table}

We also estimate the foreground power spectra in the VSA regions by
simulating observations of the $100~\mu$m map in the VSA fields based
on the actual $u,v$ positions of the visibilities and taking into
account the primary beam. The resulting power spectra are
well-modelled by a simple power-law function in each case and the
expected fall-off in power with increasing $\ell$ agreed with previous
analyses (SFD98), with power-law indices in the range $-0.5$ to
$-1.0$. Of course, the exact scaling law depends on frequency and
position on the sky. We fitted VSA data to the $100~\mu$m template for each individual VSA field. This
was difficult due to low-signal-to noise ratios and the relatively
stronger CMB anisotropies. The fitted values were in the range
$0-30~\mu$K/(MJy~sr$^{-1}$) but with large errors and no correction
for chance alignments of the template maps and the CMB. Taking the
average over all fields was very close to the canonical value
of $10~\mu$K/(MJy~sr$^{-1}$). Since we are only interested in the
overall effect in the combined power spectrum, we chose to stick with the canonical value. It was used to re-scale the power spectra fits to
each region and hence provide a statistical correction to the final
VSA power spectrum in section~\ref{sec:results_ps}. The scatter on the
derived coupling coefficients both here and from other experiments, at
similar frequencies and angular scales, suggests that the overall
scaling could be wrong by up to a factor of $\sim 2$ at most. However, the
correction is very small for these data and hence such an error would have only a minor
effect at the lowest $\ell$-values ($\ell \ltsim 100$) where the VSA
has little or no sensitivity; this is discussed further in
section~\ref{sec:discussion}. The statistical correction applied to
the final power spectrum has the form $\Delta {\rm T}^{2}_{\rm Gal} =
4.1 \times 10^{3} \times \ell^{-0.72}$. This corresponds to $28
~\mu$K$^{2}$ at $\ell = 1000$. A more detailed analysis of Galactic
foregrounds will be described in a forthcoming paper (Dickinson et al., in prep.).

\subsection{Galaxy clusters foreground $-$ the SZ effect}
\label{sec:sz_contamination}

Galaxy clusters can be a serious contaminant for CMB
observations. Large reservoirs of hot gas ($T \sim 10^{7}$~K) in the
intracluster medium re-scatter the CMB photons by inverse Compton
scattering. Since the optical depth is typically $\sim 1$~per cent,
this creates a small distortion of the CMB frequency spectrum~$-$~ the
Sunyaev-Zeldovich Effect (Zeldovich \& Sunyaev 1969; Sunyaev \&
Zeldovich 1970), henceforth SZE. For frequencies below the CMB
spectrum peak at 217~GHz, the SZE corresponds to a temperature
decrement along the line of sight to the cluster. The largest
clusters, at relatively low redshift ($z < 0.1$), can easily be
detected by the VSA (Rusholme 2001; Lancaster et al. 2004) and hence
it is crucial to avoid these when making CMB observations. Fortunately, the majority of nearby clusters have been
identified by optical surveys such as the Abell catalogue (Abell,
1958). The VSA regions are located away from known clusters
(section~\ref{sec:obs_schedule}).

One potential contaminant, which we have not discussed so far, is the
Poisson distributed contribution due to SZE decrements from virialized
clusters at higher redshifts. For $2000 < \ell < 4000$ this is likely
to have the same $\ell$-dependence as the point source contribution,
albeit with a lower amplitude. If normalisation $\sigma_8=0.9$ then we estimate,
using the cluster model described in Battye \& Weller (2003), a
confusion noise of $\approx 1~{\rm mJy}$~beam$^{-1}$, well below the
noise level in typical VSA maps, of $\approx 6~{\rm
mJy~beam}^{-1}$. If the excess power detected by CBI (Mason et
al.~2003; Readhead et al.~2004) is due to the SZE then $\sigma_{8}
\approx 1.1$ (Komatsu \& Seljak 2002) and we estimate $ {\rm
2.5~mJy~beam}^{-1}$ compatible with an extrapolation of the observed
power spectrum from $\ell=3000$. In either case, the SZE foreground is
not significant for current VSA data.

\section{RESULTS}
\label{sec:results}

\subsection{Maps}
\label{sec:maps}

The primary goal of the VSA is to measure the CMB power spectrum over
a wide range of angular scales. Sky maps also provide an important and
complementary view of the data. To create maps of the CMB we used a
maximum entropy method (MEM) described in Maisinger, Hobson \& Lasenby
(1997). The input data are the same binned visibility data used for
deriving the power spectrum in section~\ref{sec:results_ps}. We chose
to use the ``memuv'' option which reconstructs the Fourier modes in
the $u,v$-plane; these are then Fourier transformed to make maps. To test the robustness of the map-making software, we tried a
range of options with various parameters when producing the
maps. We found no significant differences
between these maps. The
MEM algorithm is preferred over other deconvolution algorithms since
it is optimized for extended emission by assuming a flat sky as a
prior. The CLEAN algorithm (H\"{o}gbom, 1974), on the other hand, is
optimized for unresolved sources which are treated as delta functions
convolved with the synthesized beam.

\begin{figure*}
\vbox to245mm{\vfil \includegraphics[angle=90]{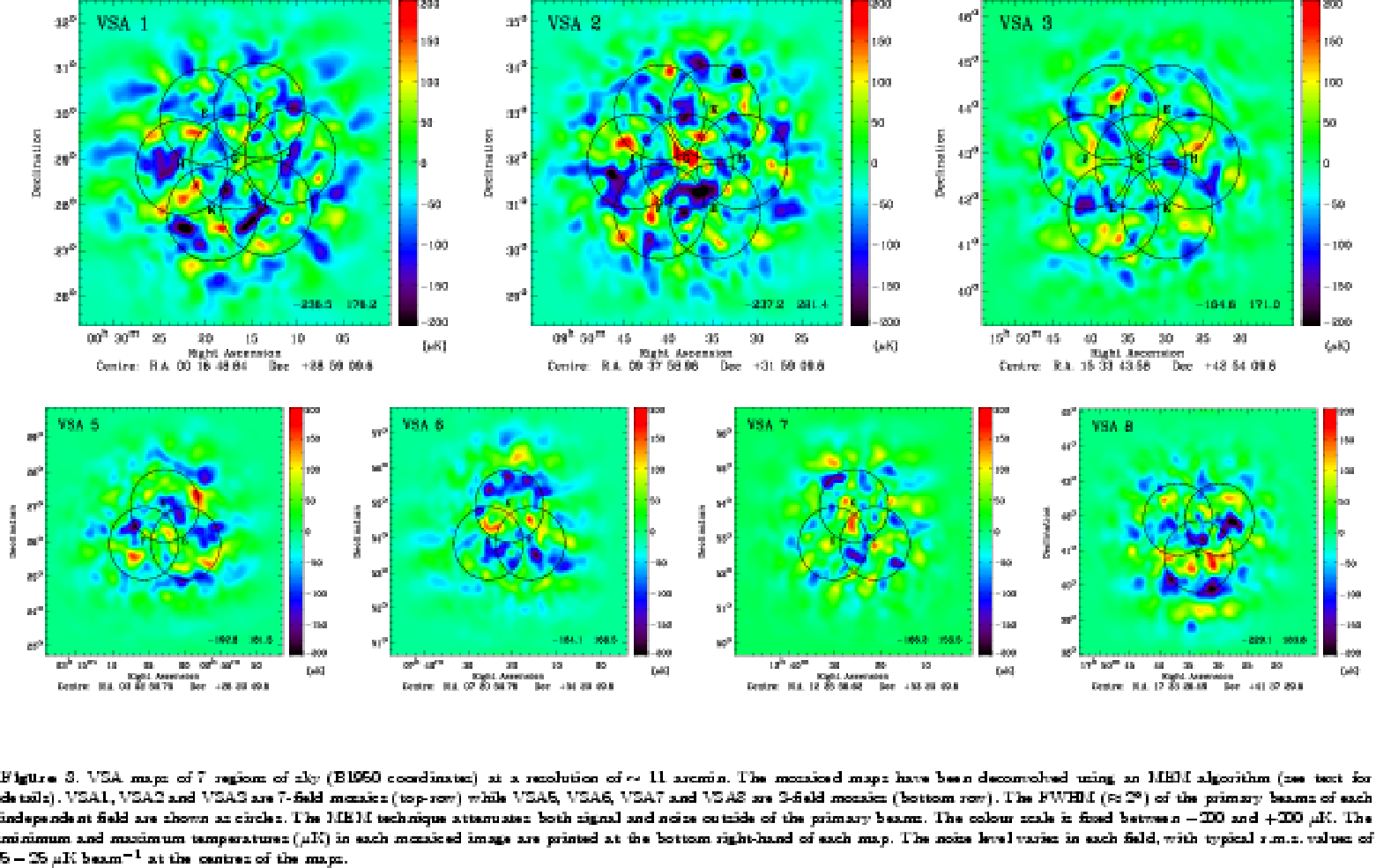} \vfil}
\caption{}
\label{fig:maps1} 
\end{figure*}

The MEM-based CMB maps made from VSA data in the extended array,
are presented in Fig.~\ref{fig:maps1}. Foreground radio sources have
been removed as explained in
section~\ref{sec:source_subtraction}. CMB anisotropies are clearly detected in all 7 VSA regions.

The primary beam envelope of an interferometer permits a separation of
the astronomical signal and the noise. Within the primary beam, the
map contains both CMB and noise components while outside the primary
beam, only noise remains. This is not the case for the MEM maps in
Fig.~\ref{fig:maps1}, so we also created maps based on the CLEAN
deconvolution algorithm with the {\sc IMAGR} task in the {\sc AIPS}
package. Table~\ref{tab:rms_noises} lists the estimates of CMB signal
and thermal noise for individual VSA fields. Significant variations in
the signal-to-noise ratio (SNR) are evident (SNR~$\sim~1-3$), primarily
due to the varying integration times after flagging and filtering of
the data. Note that the signal-to-noise ratio in the mosaiced maps (Fig.~\ref{fig:maps1}) is better due to the overlapping of the individual fields.
\begin{table}
\centering
\caption{Comparison of r.m.s. of the signal and thermal noise, all units  
mJy~beam$^{-1}$, for each VSA field, measured from the CLEANed
maps. The r.m.s. noise level in the centre of each map and the
corresponding residual CMB signal observed in each field are
given. Fields with 2 values (A and B) are for data at 2 frequencies
(34 and 33~GHz respectively).\label{tab:rms_noises}}
 \begin{tabular}{lccc}
    \hline
Field  & Thermal noise 	& r.m.s. in   	& CMB signal\\
       & (r.m.s.)      	&centre of map  &(r.m.s.) \\
       \hline
VSA1E   &8.0&18.6& 16.8\\
VSA1F   &6.9&13.8& 12.0\\
VSA1G   &8.4&11.2& 7.4 \\
VSA1H\_A&9.7&13.8& 9.8 \\
VSA1H\_B&7.9&14.7& 12.4\\
VSA1J   &5.2&13.5& 12.5\\
VSA1K   &6.0&20.1& 19.2\\
VSA1L   &4.2&11.6& 10.8\\
VSA2E   &8.9&16.5& 13.9\\
VSA2F   &9.6&18.2& 15.5\\
VSA2G   &7.1&22.5& 21.3\\
VSA2H   &4.8&11.9& 10.9\\
VSA2J\_A&11.0&18.3&14.6\\
VSA2J\_B&7.8&18.5& 16.8\\
VSA2K   &6.6&17.3& 16.0\\
VSA2L   &6.2&12.6& 11.0\\
VSA3E   &7.8&15.9& 13.8\\
VSA3F   &7.7&18.4& 16.7\\
VSA3G   &6.5&12.4& 10.5\\
VSA3H\_A&8.7&18.9& 16.8\\
VSA3H\_B&9.2&14.2& 10.8\\
VSA3J   &6.8&9.0&  5.9 \\
VSA3K   &6.9&12.3& 10.2\\
VSA3L   &5.1&11.4& 10.2\\
VSA5E   &6.7&9.0&  6.0 \\
VSA5F   &8.0&16.4& 14.3\\
VSA5G   &5.9&13.0& 11.6\\
VSA6E\_A&10.6&10.7&1.5 \\
VSA6E\_B&8.6&14.0& 11.0\\
VSA6F   &6.7&16.4& 15.0\\
VSA6G   &7.0&9.2&  6.0 \\
VSA7E\_A&8.3&7.0&  -   \\
VSA7E\_B&9.2&9.7&  3.1 \\
VSA7F   &5.4&14.1& 13.0\\
VSA7G   &6.1&13.8& 12.4\\
VSA8E\_A&5.9&14.2& 12.9\\
VSA8E\_B&16.0&13.5&-   \\
VSA8F   &3.8&7.3&  6.2 \\
VSA8G   &5.3&14.8& 13.8\\
  \hline
 \end{tabular}
\end{table}

\subsection{Power spectrum}
\label{sec:results_ps}

The final visibility data are binned into square $u,v$ cells, each 9
wavelengths on a side, to oversample the data while reducing the number of
data points by a factor of $\gtsim 1000$. Sources are
subtracted using position and flux density information as described in
section~\ref{sec:source_subtraction}. This results
in 33 visibility files, one for each VSA pointing, each with $\sim 10^{3}$
data points.

The binned visibilities form the basic input to the maximum likelihood
analysis for the CMB power spectrum. We used the Microwave Anisotropy
Dataset Computational sOftWare (MADCOW) described by Hobson \& Maisinger
(2002). 

\begin{figure}
\begin{center}
\includegraphics[width=0.5\textwidth,angle=0,scale=1.0]{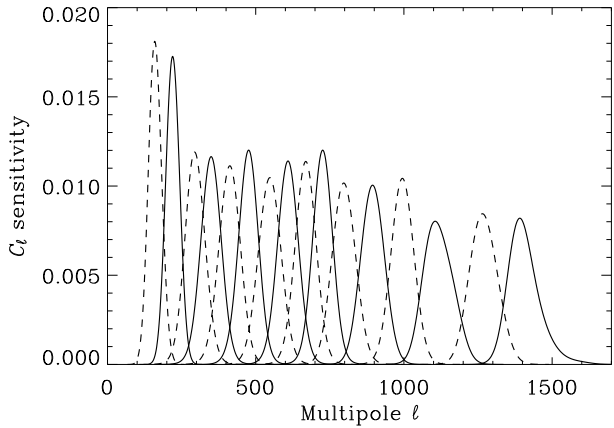}
  \caption{VSA window functions plotted as a function of $\ell$ for
  the main binning. Each
  curve is a single bin out of a total of 16 bins. Curves are solid
  and dashed line for clarity. The window
  functions are, in general, well-modelled by Gaussian functions. \label{fig:window_functions}}
\end{center}
\end{figure}

The main binning of the VSA data reduction was chosen to Nyquist
sample and to approximately match the expected phase of the peaks and troughs in the power
spectrum as closely as possible. This means that there is a bin
corresponding to the $\ell$-value expected for each peak in the power
spectrum and an additional bin between the peaks. Note that this
approach does not bias the parameter estimation towards the
concordance model (see for example, Kuo et al.~2004) $-$ it merely acts
as a matched filter that tries to extract maximum information from the
fluctuations in the power spectrum. The bin positions and widths are the same as those used in Grainge et al. (2003). The considerably larger
contiguous sky area in the seven field mosaics results in reduced
correlation between adjacent bins. The inverse correlation matrix for
the combined VSA data set is given in
Table~\ref{tab:bin_correlations}. The window functions are calculated
as described in Paper~III and are plotted in
Fig.~\ref{fig:window_functions}; they are well-fitted by Gaussian
functions. In principle, finer binning is possible but at the cost of
increasing the correlations and reducing the signal-to-noise ratio.
\begin{table}
\caption{The correlation matrix ${\cal C}_{i,j}$ for the combined VSA
data (main binning only), normalised to $1.0$. The values of the matrix for which ${\cal
C}_{i,j}$ is not reported, can be assumed 
to be zero. The final column gives the diagonal elements of the covariance
matrix in units of $10^5 \times \mu \rm{K}^4$.
\label{tab:bin_correlations}}
\begin{tabular}{ccccccc} \hline
$B$ & ${\cal C}_{B,B-2}$ & ${\cal C}_{B,B-1}$ & ${\cal C}_{B,B}$ &
${\cal C}_{B,B+1}$ & ${\cal C}_{B,B+2}$ & $\mbox{Cov}_{B,B} $\\
\hline
$ 1 $ &  &    &   $ 1.0 $  &   $ -0.165 $  &   $ 0.022 $ & $ 15.431 $ \\
$ 2 $ &  &   $ -0.165 $  &   $ 1.0 $  &   $ -0.255 $  &   $ 0.008 $ & $ 12.82 $ \\
$ 3 $ & $ 0.022 $  &   $ -0.255 $  &   $ 1.0 $  &   $ -0.126 $  &   $ 0.001 $ & $ 8.024 $ \\
$ 4 $ & $ 0.008 $  &   $ -0.126 $  &   $ 1.0 $  &   $ -0.198 $  &   $ 0.02 $ & $ 1.786 $ \\
$ 5 $ & $ 0.001 $  &   $ -0.198 $  &   $ 1.0 $  &   $ -0.209 $  &   $ 0.039 $ & $ 0.573 $ \\
$ 6 $ & $ 0.02 $  &   $ -0.209 $  &   $ 1.0 $  &   $ -0.251 $  &   $ 0.074 $ & $ 1.687 $ \\
$ 7 $ & $ 0.039 $  &   $ -0.251 $  &   $ 1.0 $  &   $ -0.324 $  &   $ 0.084 $ & $ 1.244 $ \\
$ 8 $ & $ 0.074 $  &   $ -0.324 $  &   $ 1.0 $  &   $ -0.315 $  &   $ 0.11 $ & $ 1.355 $ \\
$ 9 $ & $ 0.084 $  &   $ -0.315 $  &   $ 1.0 $  &   $ -0.402 $  &   $ 0.09 $ & $ 1.282 $ \\
$ 10 $ & $ 0.11 $  &   $ -0.402 $  &   $ 1.0 $  &   $ -0.275 $  &   $ 0.036 $ & $ 1.879 $ \\
$ 11 $ & $ 0.09 $  &   $ -0.275 $  &   $ 1.0 $  &   $ -0.133 $  &   $ 0.023 $ & $ 0.854 $ \\
$ 12 $ & $ 0.036 $  &   $ -0.133 $  &   $ 1.0 $  &   $ -0.187 $  &   $ 0.026 $ & $ 0.846 $ \\
$ 13 $ & $ 0.023 $  &   $ -0.187 $  &   $ 1.0 $  &   $ -0.173 $  &   $ 0.019 $ & $ 0.79 $ \\
$ 14 $ & $ 0.026 $  &   $ -0.173 $  &   $ 1.0 $  &   $ -0.117 $  &   $ 0.018 $ & $ 0.706 $ \\
$ 15 $ & $ 0.019 $  &   $ -0.117 $  &   $ 1.0 $  &   $ -0.157 $  &   & $ 1.255 $ \\
$ 16 $ & $ 0.018 $  &   $ -0.157 $  &   $ 1.0 $  &    &   & $ 4.261 $ \\
\hline
\end{tabular}
\end{table}

\begin{figure*}
\begin{center}
\includegraphics[width=0.5\textwidth,angle=90,scale=0.7]{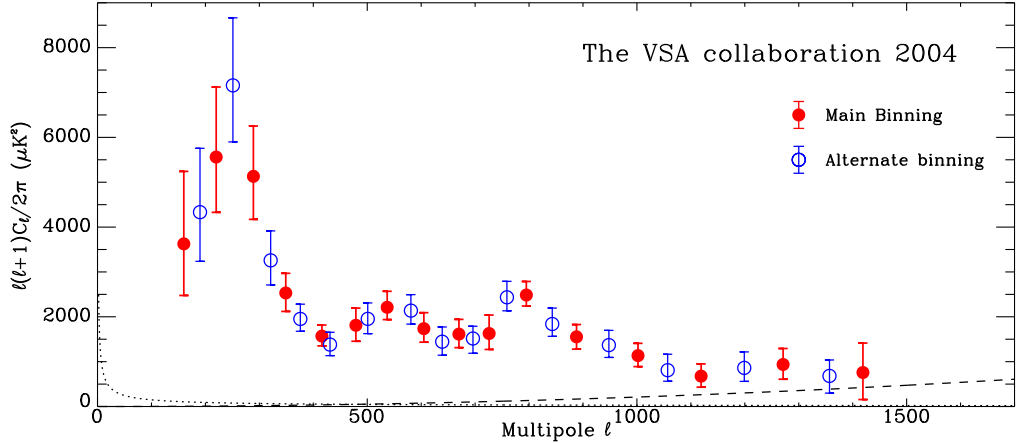}
  \caption{The CMB power spectrum as measured by the VSA by combining
  the extended and compact array data for all 7 VSA regions. The errors represent $1\sigma$
  limits from both thermal noise and sample variance. Two alternate binnings (filled red circles and unfilled blue circles) are displayed
  (Table~\ref{tab:bandpowers}). Absolute calibration is accurate to
  3~per cent and is not included in the errors. The bright radio sources have been subtracted. The smaller corrections for residual sources and Galactic foregrounds have been made and are shown as a dashed line and dotted line respectively. \label{fig:ps}}
\end{center} 
\end{figure*}

The final band powers calculated from the complete VSA data set,
including both the compact and extended array data sets, are listed in
Table~\ref{tab:bandpowers} for two independent binnings\footnote{The
VSA data are available at the following URL: \\ {\tt
http://www.jb.man.ac.uk/research/vsa/vsa\_results.html}}. The final
combined VSA power spectrum is presented in Fig.~\ref{fig:ps}. Bright
foreground sources (section~\ref{sec:source_subtraction}) have been
subtracted from the visibility data before power spectrum
estimation. The small corrections for residual sources
(section~\ref{sec:source_results}) and diffuse foregrounds
(section~\ref{sec:diffuse_foreground_estimates}) have been applied and
are also plotted in Fig.~\ref{fig:ps}. The error bars were calculated
from the probability likelihood functions by enclosing $68$ per cent
of the area centred on $\ell_{h}$, the weighted median $\ell$ value
for each bin. Calibration uncertainty (3 per cent; section~\ref{sec:abs_cal}) is not included. Sample variance is
included in the error estimates. Note that the extended array data
have little sensitivity at $\ell \ltsim 300$ and hence the first 3
bins come mainly from data taken in the compact array described in
Paper~III.

\begin{figure*}
\begin{center}
\includegraphics[width=0.5\textwidth,angle=90,scale=0.7]{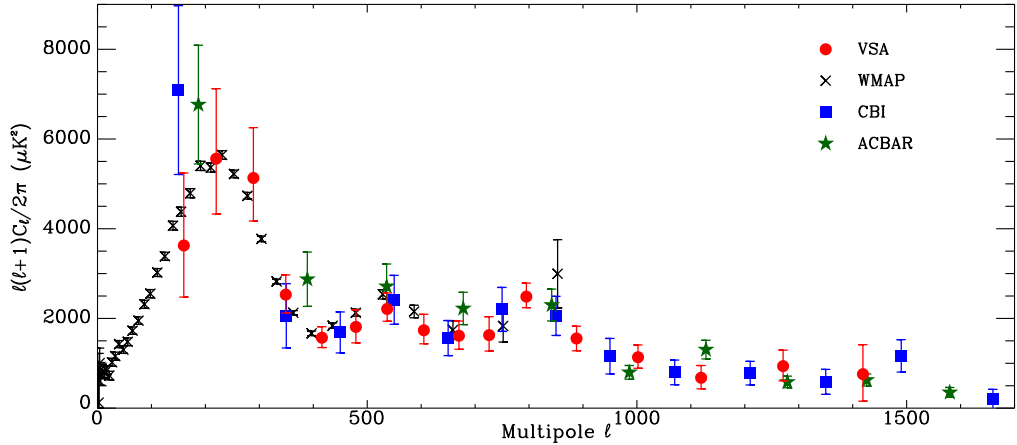}
\caption{A comparison of the VSA data (red filled circles) with those
  from WMAP 1-year (crosses), CBI (filled blue squares) and ACBAR
  (filled stars). All errors are at the $1\sigma$ level. There is good
  agreement between the different experiments. \label{fig:ps_compare}}
  \end{center}
\end{figure*}

The VSA data are also plotted in Fig.~\ref{fig:ps_compare} along with
recent data from other CMB experiments; WMAP (Bennett et al. 2003a),
ACBAR (Kuo et al. 2004) and CBI (Readhead et al. 2004). Only a single
binning of the data shown. No corrections are made to the external
measurements for the binning schemes; they are plotted
directly. Nevertheless, the overall consistency between these
experiments is remarkable.
\begin{table}
\caption{The CMB band-powers (in $\mu$K$^{2}$) for main and offset binnings of the complete
VSA data set combining both compact and extended array data. The
quoted $\ell$-ranges are nominal bin limits when the window functions
fall to half the peak value, while $\ell_h$ is the
weighted median value of the relevant window function. The error bars enclose
68 percent of the likelihood and include sample variance. \label{tab:bandpowers}}
\centering
\begin{tabular}{ccccc}
\hline
Bin & $\ell$-range & $ \ell_{h}$ &  $T_0^2\ell(\ell+1)C_{\ell}/2\pi~ [\mu
{\rm K}^2]$ \\
\hline
1 & $100 - 190 $ & $160$ &  $3626 _{ -1150} ^{ +1616  }$ &\\
1A & $145 -  220 $ & $190$ & &$4430 _{-1096} ^{+1424} $ \\
2 & $190 - 250 $ & $220$ &  $5561 _{ -1232} ^{ +1561  }$ &\\
2A & $220 -  280 $ & $251$ & &$7236 _{-1260} ^{+1506} $ \\
3 & $250 - 310 $ & $289$ &  $5131 _{ -959} ^{ +1123  }$ &\\
3A & $280 -  340 $ & $321$ & &$3324 _{-548} ^{+657} $ \\
4 & $310 - 370 $ & $349$ &  $2531 _{ -411} ^{   +438  }$ &\\
4A & $340 -  410 $ & $376$ & &$2010 _{-274} ^{+329} $ \\
5 & $370 - 450 $ & $416$ &  $1570 _{ -219} ^{   +246  }$ &\\
5A & $410 -  475 $ & $431$ & &$1432 _{-246} ^{+274} $ \\
6 & $450 - 500 $ & $479$ &  $1811 _{ -356} ^{   +383  }$ &\\
6A & $475 -  540 $ & $501$ & &$2000 _{-329} ^{+356} $ \\
7 & $500 - 580 $ & $537$ &  $2212 _{ -274} ^{   +356  }$ &\\
7A & $540 -  610 $ & $581$ & &$2180 _{-301} ^{+356} $ \\
8 & $580 - 640 $ & $605$ &  $1736 _{ -301} ^{   +356 }$ &\\
8A & $610 -  670 $ & $639$ & &$1484 _{-301} ^{+329} $ \\
9 & $640 - 700 $ & $670$ &  $1614 _{ -301} ^{   +329  }$ &\\
9A & $670 -  725 $ & $696$ & &$1553 _{-329} ^{+356} $ \\
10 & $700 - 750 $ & $726$ &  $1628 _{ -356} ^{   +411  }$& \\
10A & $725 -  800 $ & $759$ & &$2469 _{-301} ^{+356} $ \\
11 & $750 - 850 $ & $795$ &  $2486 _{ -246} ^{   +301  }$& \\
11A & $800 -  900 $ & $843$ & &$1871 _{-274} ^{+274} $ \\
12 & $850 - 950 $ & $888$ &  $1553 _{ -274} ^{   +274  }$& \\
12A & $900 - 1000$ & $948$ & &$1398 _{-274} ^{+274} $ \\
13 & $950 - 1050$ & $1002$ & $1135 _{-246} ^{  +274  }$ &\\
13A & $1000 - 1125$ & $1057$ & &$837 _{-246} ^{+246} $ \\
14 & $1050 - 1200$ & $1119$ & $677 _{ -246} ^{   +274  }$& \\
14A & $1125 - 1275$ & $1199$ & &$886 _{-301} ^{+329} $ \\
15 & $1200 - 1350$ & $1271$ & $937 _{ -329} ^{  +356  }$ &\\
15A & $1275 - 1525$ & $1357$ & &$704 _{-383} ^{+329} $ \\
16 & $1350 - 1700$ & $1419$ & $758 _{ -603} ^{  +657  }$&\\
\hline
\end{tabular}
\end{table}

\subsection{Morphological power spectrum analysis}
\label{sec:ps_gaussians}

The shape of the power spectrum was analysed following the technique described in \"Odman et al. (2003). By fitting
multiple Gaussians, it calculates the best-fit to the data. Each
Gaussian is defined by its amplitude in temperature $\Delta T$, centre
multipole $\ell$ and width $\sigma$.

The data were split into subsets according to the nature of each
experiment. The low-frequency (LF) subset consists of  DASI (Halverson
et al. 2002), CBI (Mason et al. 2003; Pearson et al. 2003) and VSA
data (this work), with frequencies $\sim 30$~GHz. The high frequency
(HF) subset, corresponding to frequencies of $\sim 100$~GHz,
comprises Archeops (Beno{\^i}t et al. 2003), ACBAR (Kuo et al. 2004),
BOOMERanG (Netterfield et al. 2002) and MAXIMA (Lee et al. 2001)
data. An additional analysis on the VSA data alone was carried out to
check for consistency. WMAP data were not included to simplify the low
and high frequency split.

The results of this analysis $-$ the amplitudes and centre $\ell$-values
for each peak in the spectrum $-$ are in
Table~\ref{tab:gauss_fits}. LF data do not tightly constrain the
first peak position owing to the lack of data points at low
$\ell$. This appears as a double peak in the likelihood for $\ell_1$
and $\Delta T_1$. The uncertainty on the first peak does not propagate
at higher multipoles where the secondary peaks are increasingly
well-defined. HF data provide constraints on the 1st, 2nd and 3rd
peaks whereas LF data provide constraints on the 2nd to the 5th peaks.
An important aspect of this analysis is the apparent discrepancy, at
$\approx 2 \sigma$ level, between LF and HF data on the amplitude of
the 3rd peak. The LF data prefer a slightly higher 3rd peak of
$\approx 48~\mu$K, compared to the HF data which prefer a lower value
of $\approx 40~\mu$K. This feature has been noted before and mostly accounts for the high baryon fraction derived from LF data alone.

This type of model-dependent analysis is often difficult to
interpret. Specifically, the 4th and 5th peaks are not real
detections and are merely a consequence of how the fitting procedure
accounts for the exponential drop in power due to the damping
tail. Nonetheless, the structure of the CMB power spectrum, at least
at $\ell < 1000$, is relatively well-defined. There is good
agreement between this analysis and a similar one made for the 1-year
WMAP data on the 1st and 2nd peaks (Page et al. 2003).
\begin{table*}
\centering
\caption{Results of fitting Gaussians to the CMB power spectra as
measured by various CMB experiments. Here we show four combinations of
data (see text), including low frequency (LF; $\nu \sim 30$~GHz) and
high frequency (HF; $\nu \sim 100$~GHz) data. Amplitudes are in units
of $\mu$K. Errors are at the 68 per
cent confidence level; ``nc'' is where there are no constraints.}
\begin{tabular}{|l|c|c|c|c|}\hline
\label{tab:gauss_fits}
parameter & Full data set & High Frequency & Low Frequency & VSA only
\\ \hline $\Delta T_1$ & $71.1^{+2.7}_{-1.9}$ & $67.1^{+4.9}_{-2.3}$ &
$75.5^{+1.7}_{-4.7}$ & $83.0^{+7.4}_{-3.4}$ \\ $\Delta T_2$ &
$43.9^{+2.0}_{-0.3}$ & $42.2^{+3.0}_{-5.4}$ & $43.9^{+1.4}_{-1.0}$ &
$47.4^{+3.6}_{-3.2}$ \\ $\Delta T_3$ & $48.0^{+1.8}_{-1.0}$ &
$40.2^{+3.1}_{-3.5}$ & $48.1^{+0.9}_{-3.1}$ & $51.2^{+3.7}_{-6.5}$ \\
$\Delta T_4$ & $32.3^{+0.6}_{-2.3}$ & $32.0^{+7.5}_{-5.8}$ &
$32.0^{+1.1}_{-1.7}$ & $34.9^{+2.4}_{\mbox{{\tiny nc}}}$ \\ $\Delta
T_5$ & $29.6^{+2.8}_{-0.8}$ & $25.2^{+7.2}_{-7.2}$ &
$29.0^{+2.0}_{-0.3}$ & $36.8^{+2.4}_{-2.3}$ \\ $\ell_1$ &
$217^{+2}_{-4}$ & $214^{+4}_{-11}$ & $189<\ell_1<236$ &
$241^{+1}_{-8}$ \\ $\ell_2$ & $552^{+3}_{-6}$ & $513^{+25}_{-20}$ &
$546^{+9}_{-3}$ & $538^{+11}_{-5}$ \\ $\ell_3$ & $808^{+2}_{-18}$ &
$814^{+43}_{-60}$ & $810^{\mbox{{\tiny nc}}}_{-19}$ &
$795^{+21}_{-11}$ \\ $\ell_4$ & $1001^{+15}_{-54}$ &
$1132^{+54}_{-78}$ & $969^{\mbox{{\tiny nc}}}_{-41}$ &
$988^{+31}_{-38}$ \\ $\ell_5$ & $1322^{+12}_{-15}$ &
$1469^{+92}_{-135}$ & $1302^{+16}_{-21}$ & $1323^{+11}_{-22}$ \\
\hline
\end{tabular}
\end{table*}

\section{DISCUSSION}
\label{sec:discussion}

\subsection{Maps}
\label{sec:discuss_maps}

An important application of CMB maps is to look for the signature of
any non-Gaussian structures which may be foregrounds or primordial in
origin, such as cosmic strings which would appear as a discontinuity
on a map, or a SZE which would produce a negative temperature against
the CMB.

\begin{figure*}
\begin{center}
\includegraphics[width=1.0\textwidth,angle=0,scale=0.8]{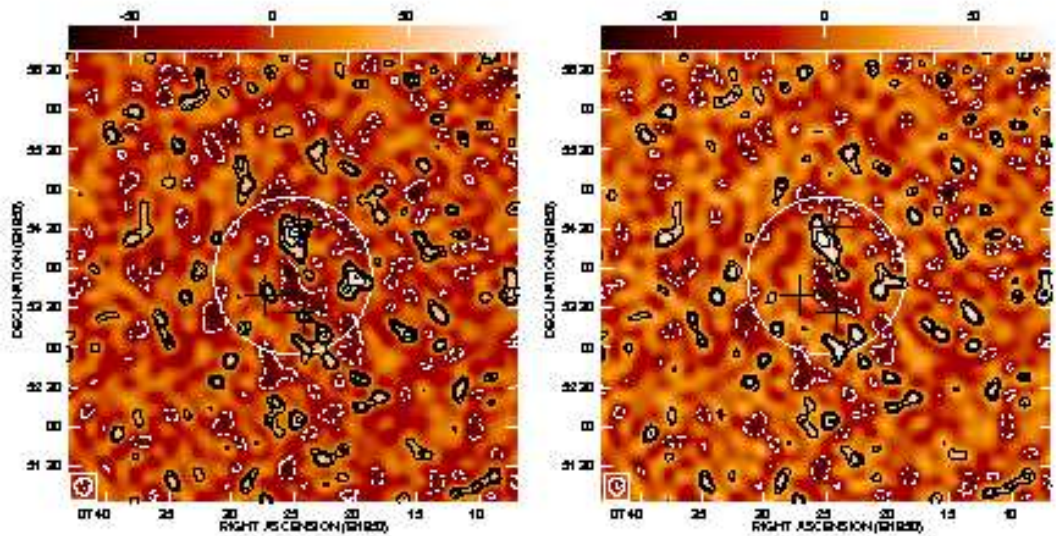}
  \caption{VSA maps of VSA6E field before and after source
  subtraction (Cleary 2003). The maps have been deconvolved using the
  CLEAN
  algorithm. The FWHM of the primary beam is indicated by the
  circle. Astronomical signals are inside the primary beam while the
  edges of the map, outside the primary beam, contain noise. The crosses indicate the positions of the three brightest
  sources in this field (see text). The r.m.s. noise level is $\approx 10$~mJy.\label{fig:sources_compare} } 
\end{center}
\end{figure*}

The largest foreground component for the VSA is discrete sources. The
stronger sources can be identified in the raw VSA maps, before source
subtraction. They show up as large positive temperatures in the maps,
although they can coincide with negative CMB features (or noise),
making faint sources difficult to identify in the sky maps. A
comparison of an un-subtracted map  with one that has had sources
removed is presented in Fig.~\ref{fig:sources_compare}. These maps
have been deconvolved with the CLEAN algorithm to optimise detection of discrete sources. In VSA6E, there are 3 relatively strong sources,
even after correction for the primary beam attenuation, with apparent
flux densities of 91, 84 and 39~mJy. The r.m.s. noise in the data is
$\approx 10$~mJy. The requirement for accurate source subtraction, as
used by the VSA, is evident. 
The VSA maps allow a comparison to be made with other CMB data. The
recent WMAP data release (Bennett et al. 2003) has provided 5 all-sky
maps at frequencies centred at 22.8 (K-band), 33 (Ka-band), 40.7
(Q-band), 60.8 (V-band) and 93.5~GHz (W-band) with resolutions ranging
from 49.2~arcmin (K-band) to 12.6 arcmin (W-band). The signal-to-noise
ratio of WMAP data at the VSA resolution is $\sim 1$ and hence much of
the CMB signal is lost in the noise. The actual noise level in the
WMAP data depends on position due to the scanning strategy of the WMAP
satellite. For the 1-year WMAP data release, the noise is $\sim
100-200~\mu$K per $12.6$~arcmin pixel in the VSA regions, compared to
$\sim 25~\mu$K in the VSA mosaiced maps.

\begin{figure*}
\begin{center}
\includegraphics[width=1.0\textwidth]{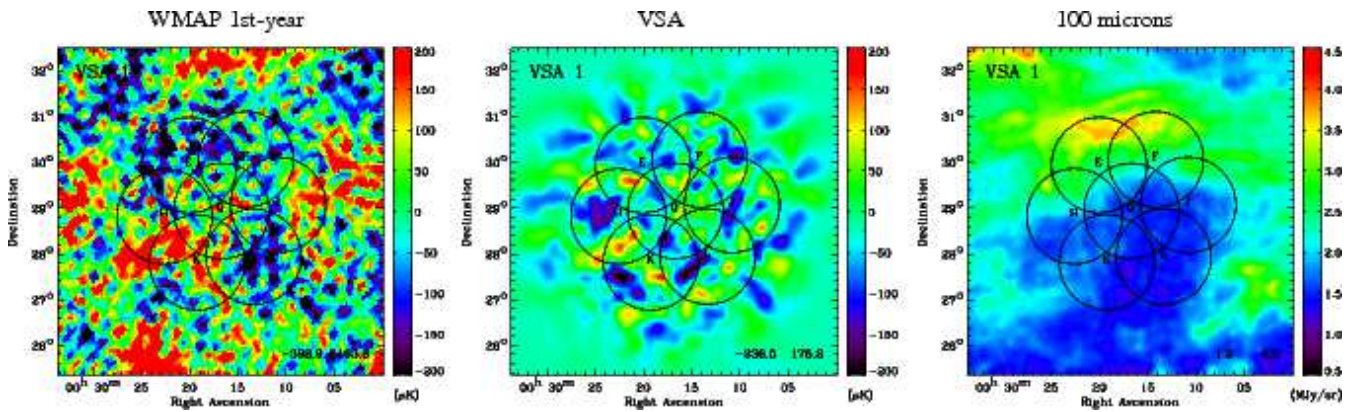}
  \caption{Comparison of maps in region VSA1. {\it Left}:
  Foreground-cleaned CMB map based on WMAP 1-year data as re-analysed by
  Tegmark et al. (2003), with a resolution of 12.6~arcmin. {\it Middle}:
  Source-subtracted CMB map from VSA data at a resolution of $\sim~11$~arcmin. {\it Right}: SFD98 $100~\mu$m map (in units of MJy~sr$^{-1}$) smoothed to
  11~arcmin. The WMAP noise level is $\sim 100~\mu$K ($1\sigma$).\label{fig:vsa1_tegmark}}
\end{center}
\end{figure*}

The WMAP team provide their own foreground-cleaned CMB maps by
combining all the WMAP channels and optimising the individual weights
to a CMB signal. However, the resolution is constrained by the K-band
channel at a resolution of 49~arcmin. More recently, Tegmark, de
Oliveira-Costa \& Hamilton (2003) have produced a foreground-cleaned
CMB map at the resolution of the highest frequency channel
(12.6~arcmin). In Fig.~\ref{fig:vsa1_tegmark}, we compare the VSA data
for region VSA1, with the foreground-cleaned CMB map of Tegmark et
al. (2003), based on WMAP 1-year data at a resolution of
$12.6$~arcmin. The colour scale uses the same range ($-200$ to
$+200~\mu$K) as that used in Fig.~\ref{fig:maps1}. Some features, both
positive and negative, can be seen on both maps. The correlation is
however restricted due to the larger noise level in the WMAP map for
this field. The minimum and maximum values for these maps are
typically $\sim \pm 400~\mu$K, while for the VSA maps the minimum and
maximum values are $\sim 250~\mu$K, in agreement with simulated CMB
maps based on the concordance model and a noise level of
$\sim 100~\mu$K. Fig.~\ref{fig:vsa1_tegmark} also shows the SFD98 $100~\mu$m dust map, smoothed
to 11~arcmin resolution. For the VSA1
region, which contains moderate dust emission, there are no obvious
Galactic signals in either the VSA or WMAP data for this region. It is
evident that the strongest Galactic emissions are on large angular
scales ($\ell \ltsim 100$), at least at high Galactic latitudes ($|b|
\gtsim 30^{\circ}$). In general, the VSA maps are consistent with each other with no obvious sign of contamination from foregrounds or other correlated signals. 
\subsection{Power spectrum}
\label{sec:discuss_ps}

The VSA power spectrum (Fig.~\ref{fig:ps}) clearly shows the existence
of the first three acoustic peaks and the exponential fall-off in
power towards higher $\ell$. The WMAP all-sky data provides
measurements limited by cosmic variance, at $\ell < 350$. At
moderate to high $\ell$ values ($\ell \gtsim 800$) there are fewer measurements of the CMB power
spectrum. Recently, the ACBAR and CBI instruments have released new
results which extend the $\ell$-range further. The morphological
analysis presented in section~\ref{sec:ps_gaussians} also suggests
very good agreement between experiments, particularly on the 1st and
2nd peaks. The slight discrepancy in the amplitude of the third peak
might suggest a change in the cosmological model. However, since the
discrepancy is between low and high frequency data, it is possible that this could be a consequence of some systematic effects either in the LF or HF data. 

An important aspect of these results is the improved absolute
calibration. We have used the WMAP brightness
temperature for Jupiter resulting in a reduction of the VSA
values by $8$ per cent and an  absolute uncertainty of $3$ per cent in the
power spectrum. Interestingly, with the new calibration scheme, the VSA data are
slightly lower in amplitude, at least over the $\ell$-range $350-700$. In fact, if
we had used the original calibration scheme, we would be in closer
agreement to the 1-year WMAP power spectrum. We
find that the best-fitting scaling factor is $\approx 1.05$ i.e. the
new VSA data are lower than the WMAP data by $1.6 \sigma$.

Foregrounds in the VSA data are expected to have a negligible effect
on the final power spectrum. The brighter sources have already been
subtracted prior to power spectrum estimation
(section~\ref{sec:source_subtraction}). The statistical corrections
for residual sources and the Galactic foreground are relatively small
as depicted in Fig.\ref{fig:ps}. Nevertheless, it is clear that the residual sources
are becoming significant in the highest $\ell$-bins. This emphasises the need for deeper source surveys if we are to accurately
measure CMB fluctuations at $\ell \gtsim 1500$. On the other hand, the
diffuse Galactic foregrounds are stronger at low $\ell$-values. In the
$\ell$-range covered here, the foreground maps in the VSA regions
suggest typical Galactic signals of a few tens of $\mu$K$^{2}$ at
$\ell = 1000$ compared to the CMB at $\sim 1000~\mu$K$^{2}$. The
exception to this is VSA5, and to a lesser extent VSA1. The location of these fields (see Fig.~\ref{fig:field_positions}) explains why the foreground estimates (Table~\ref{tab:gal_foregrounds}) are higher than in the other fields. The diffuse
foregrounds may indeed be stronger ($\sim 100~\mu$K$^{2}$) in these fields but they still are below the CMB anisotropies ($\sim 1000~\mu$K$^{2}$). This is verified  by the fact that there are no obvious foreground signals
seen either in the VSA maps or in the individual power spectra for
each field. Moreover, the relative weights of the different VSA fields
means that the combined power spectrum would still remain virtually
unaffected $-$ VSA5 contributes just 5 per cent to the total weight of the data as a whole. This is a major advantage of observing a number of independent
fields before combining them to increase the signal-to-noise ratio in
the power spectrum.

As for other systematics, the data from each individual field are
tested using a Bayesian joint power spectrum and non-Gaussianity
estimation discussed in Rocha et al. (2001).  This method operates
directly on the visibilities and, as such, is ideally suited to the
detection of any residual systematics (Savage et
al. 2004). Additionally, the method properly accounts for the presence of
noise in the data, making it extremely sensitive to any non-Gaussian
signal. However, we find no evidence of significant residual systematics in the data.



\section{CONCLUSIONS}
\label{sec:conclusions}

We have presented high sensitivity and foreground subtracted
measurements of the CMB power spectrum up to $\ell =1500$ observed with the
VSA. The cosmological interpretation is described in a companion paper
(Rebolo et al.~2004). The final extended array data set contains a
factor of $\sim 4$ more than those presented by Grainge
et al. (2003). The sky-coverage has increased by a factor of $\sim 3$
utilising 4 additional 3-field mosaics and increasing the original
3-field mosaics to 7-field mosaics, resulting in a significant
increase to the signal-to-noise ratio. Mosaicing techniques allow for
smaller correlations between adjacent bins and hence, can give finer
resolution in $\ell$. However, in this paper we have used the same binning scheme as Grainge et al. (2003) for comparison and to reduce the bin-bin correlations
; the $\ell$ resolution is $\Delta \ell \approx 60$. We adopt a new
absolute calibration scheme based on Jupiter temperature measurements
with WMAP to give a 3 per cent accurate absolute scale in the power
spectrum. This reduces absolute values by 4 per cent in temperature or
8 per cent in the power spectrum compared to previous VSA data.

The VSA power spectrum (Fig.~\ref{fig:ps}) is in good agreement with
data from other CMB experiments. There are now good signal-to-noise detections of the first
3 peaks in the CMB power spectrum and a damping tail at high $\ell$
which continues to $\ell \sim 1500$. The different instruments have
different potential systematic problems and they cover a wide range of
frequencies. The agreement therefore indicates that none of the
experiments are seriously contaminated by foreground emission or other
systematic effects (see also Griffiths \& Lineweaver (2004)). From the analysis of the peak structure (section~\ref{sec:ps_gaussians}), there is
extremely good agreement on the peak structure between experiments,
except for a slight discrepancy in the amplitude of the 3rd peak which
may have important consequences for the cosmological models. The slight discrepancy ($\sim 1.6\sigma$) in the overall scaling between VSA and WMAP may also be an important factor in terms of fitting cosmological models. This is discussed by Rebolo et al. (2004). 

In the near future, the VSA will be reconfigured with even larger
horns and longer baselines. It will allow the VSA to increase the
$\ell$-range even further, up to a maximum of $\ell \sim 3000$, while
maintaining good $\ell$ resolution without loss of overall temperature
sensitivity. The assistance of high frequency radio source surveys, to identify and measure the flux densities of
sources over large areas of sky, will be a crucial role in keeping
the discrete source foregrounds under control. The One Centimetre
Receiver Array (OCRA; Browne et al. 2000), a prototype multi-beam
receiver system currently begin tested on the Torun 32~m telescope,
will be ideal for such purposes. Moreover, OCRA is currently operating
at $30$~GHz and hence direct source subtraction, without the need to
extrapolate to other frequencies will be possible in the near future. 


\section*{ACKNOWLEDGMENTS}

First of all we thank the anonymous referee for detailed comments
which have significantly improved the style of the paper and presentation of the results. CD is grateful to Tim Pearson for reading the text carefully and making useful comments on the paper. We
thank the staff of Jodrell Bank Observatory, Mullard Radio Astronomy
Observatory and the Teide Observatory for assistance in the day-to-day
operation of the VSA. We are very grateful to PPARC for the funding and
support for the VSA project and to the IAC for supporting and
maintaining the VSA in Tenerife. Partial financial
support was provided by Spanish Ministry of Science and Technology
project AYA2001-1657. JARM acknowledges the hospitality of  the IAC
during several visits. CD acknowledges PPARC for funding a
post-doctoral research associate position for part of this work. KL, NR, AS and RS acknowledge are funded by PPARC studentships. YAH is supported by the
Space Research Institute of KACST. AS acknowledges the support of
St. Johns College, Cambridge. CO thanks the African Institute for
Mathematic Sciences for their hospitality.


\bibliographystyle{mn2e}


\bsp 

\label{lastpage}

\end{document}